\documentclass[11pt]{article}
\usepackage{moriond,epsfig}

\def\be{\begin{equation}}
\def\ee{\end{equation}}
\def\bea{\begin{eqnarray}}
\def\eea{\end{eqnarray}}

\begin{document}
\vspace*{4cm}
\title{The Origin of the IMF from Core Mass Functions}

\author{S.P. Goodwin, A.P. Whitworth \& D. Ward-Thompson}

\address{Department of Physics \& Astronomy, Cardiff University \\
5 The Parade, Cardiff, CF24 3YB, UK}

\maketitle\abstracts{
We examine the initial mass functions (IMFs) of stars produced by different
molecular core mass functions.  Simulations suggest that more massive
cores produce more stars, so we propose a model in which the average 
number of stars formed in a core is equal to the initial number of
Jeans masses in that core.  Small-$N$ systems decay through dynamical
interactions, ejecting low-mass stars and brown dwarfs which populate
the low-mass tail of the IMF.  Stars which remain in
cores are able to competitively accrete more gas and become more
massive.  We deduce the forms of the core mass functions required to
explain the IMFs of Taurus, Orion, IC 348 and NGC 2547.  These core
mass functions fall into two categories - one which peaks at a few
$M_\odot$ to explain Taurus and NGC 2547, and one that peaks at around
$0.2 M_\odot$ to explain Orion and IC 348.}


\section{Introduction}

All stars form in dense molecular cores (eg. Andr\'e et al. 2000).
Observations of the densest cores, known as prestellar cores
(Ward-Thompson et al. 1994), show that their mass functions are
remarkably similar to the IMF of field stars (Motte et al. 1998; Testi
\& Sargent 1998; Motte et al. 2001).  This suggests that
the form of the IMF may be directly related to the form of the 
core mass function (CMF).

Most stars $>1 M_\odot$ exist in binary or multiple systems (eg. 
Duquennoy \& Mayor
1991).   Most of these multiple systems {\em must} form
as such, since it has been shown that dynamical evolution is
unable to significantly alter the initial binary properties or
population (Kroupa 1995).  Therefore many cores must produce
multiple objects and so the IMF cannot be a simple mapping of the CMF.

In a previous paper (Goodwin et al. 2004c) we showed that the IMF of
Taurus could be explained if all of the stars in Taurus formed from
cores of a few solar masses, a CMF similar to that observed by
Onishi et al. (2002).  In this contribution we investigate the 
effect of changing the CMF on the IMF and multiplicity of star 
forming regions.

\section{Multiple star formation in cores}

Recent studies have shown that within massive ($>5 M_\odot$)
turbulent cores, multiple star formation is the norm (Bate et
al. 2002, 2003; Delgado Donate et al. 2004; Goodwin et
al. 2004a,b).  A significant population of low-mass stars and brown dwarfs
is formed by ejections from unstable multiple systems in these cores.

\begin{figure}
\centerline{\psfig{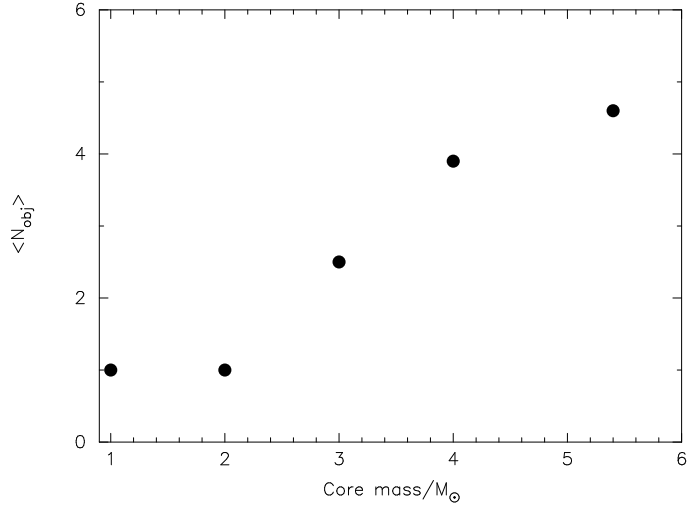}}
\caption{The average number of objects $<N_{\rm obj}>$ formed in cores
of different masses, each with the same initial turbulent virial ratio
$\alpha_{\rm turb} = E_{\rm turb}/|\Omega| = 0.1$.  The average number of objects scales roughly
linearly with the initial number of Jeans masses in the core.  }
\label{fig:ncores}
\end{figure}

Delgado Donate et al. (2003) modelled the origin of the IMF by 
assuming that fragmentation in cores is scale-free: ie. that all cores 
produce the same number of objects (stars and brown dwarfs) and that 
the masses of these objects scale with the mass of the core.  By
convolving the outcome of star formation in a $1 M_\odot$ core with a
core mass function (CMF) they obtained an IMF.  However, Goodwin et al. (in 
preparation) find that the number of objects that form depends strongly upon
the mass of a core, with low-mass cores being far less able to form
multiple objects than more massive cores.  Fig.~\ref{fig:ncores} shows 
the average number of objects that form in
cores of different masses but with the same initial thermal and turbulent
virial ratios (the ratio of the initial thermal or turbulent energy to
the initial potential energy).  The average number of objects that form is
approximately one per initial Jeans mass ($\sim 1 M_\odot$).

Given that most stars form in multiple systems and the number of
stars forming in a core might be expected to increase with the mass 
of the core, we propose a simple model for the formation of stars 
within cores and the relationship of stellar masses and 
multiplicities to the core mass (the core-to-star relationship):

\begin{itemize}

\item{Cores form an average number of objects (stars and/or brown dwarfs) 
approximately equal to the initial number of Jeans masses in the core 
(eg. Goodwin et al. in prep).}

\item{Multiple systems with $\geq 3$ members are initial unstable and
will decay to a stable system within a few $\times 10^4$ yrs 
through the ejection of low-mass stars and brown dwarfs (cf. Reipurth
\& Clarke 2001; Bate et al. 2002; Sterzik \& Durisen 2003; 
Goodwin et al. 2004a).}

\end{itemize}

The initial mass function (IMF) is then due to the convolution 
of the core mass function (CMF) {\em and} the core-to-star
relationships in these different cores.  

We assume a form for the CMF - in this case a log-normal which may have
different variances above and below the mean - and randomly sample
cores from that CMF.  A core then produces $N_*$ objects where $N_*$
is drawn from a gaussian of mean $M_{\rm core}$ (where $M_{\rm core}$ is the
core mass in solar masses) with $\sigma = 2$.  $N_*$ is then rounded
to the nearest integer $\geq 1$.

If $N_* \leq 3$ then $N_*$ stars are formed of mean mass
$\epsilon M_{\rm core}/N_*$ (we assume that the core-to-star efficiency 
$\epsilon = 0.75$ in all cases).  

If $N_* > 3$ then $N_* - 3$ stars are ejected with masses drawn
uniformly from a logarithmic distribution between $0.02 M_\odot$ and
$0.1 \epsilon M_{\rm core}$.  The remaining three stars then distribute the
rest of the mass in the core between themselves such that their
individual masses are $\epsilon(M_{\rm core} - M_{\rm ej})/3$ (where 
$M_{\rm ej}$ is the mass of ejected stars).

\section{Results}

\subsection{The IMFs of Taurus and NGC 2547}

\begin{figure}
\centerline{\psfig{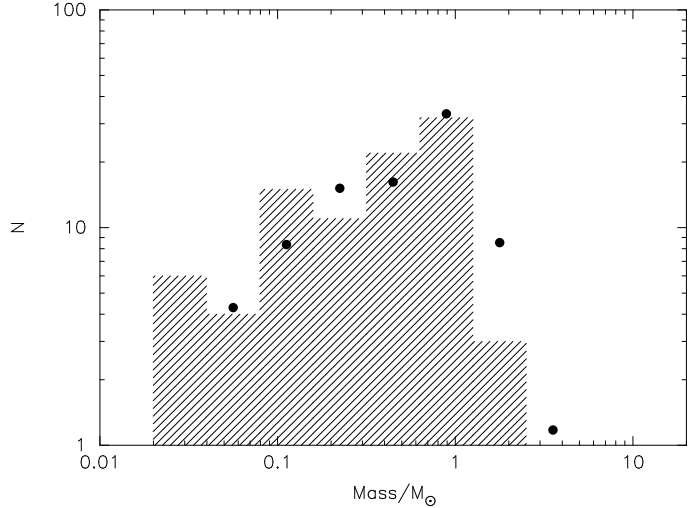}}
\caption{The IMFs of Taurus (histogram, from Luhman et al. 2003a) and
NGC 2547 (points, from Jeffries et al. 2004).  NGC 2547 has been
normalised to contain the same total number of stars as Taurus for
ease of comparison.}
\label{fig:solar-type}
\end{figure}

Both Taurus (Luhman et al. 2003a) and NGC 2547 (Jeffries et al. 2004)
have similar MFs.  Both of these MFs show a
significant peak at $\sim 1 M_\odot$, with a rapid drop above this peak,
and a rather flatter decline into the brown
dwarf regime, as illustrated in Fig.~\ref{fig:solar-type} (where the MF
of NGC 2547 has been normalised to have the same total number of stars
as Taurus for ease of comparison).  The similarity between the two 
MFs is clear.

\begin{figure}
\centerline{\psfig{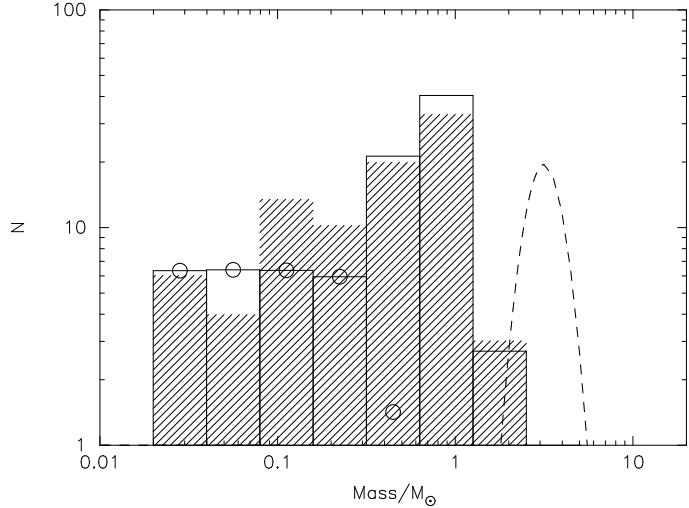}}
\caption{The open histogram shows the IMF resulting from the CMF shown
by the dashed line.  Open circles show the contribution to each bin of
ejected stars and brown dwarfs.  The hashed histogram shows the Luhman
et al. (2003a) Taurus IMF.  Both histograms are normalised to contain
the same number of stars.}
\label{fig:TAimf}
\end{figure}

Fig~\ref{fig:TAimf} shows the results of applying our
model to a log-normal CMF of mean $\bar{ {\rm log} M_{\rm core} } =
0.5$ and $\sigma_{ {\rm log} M_{\rm core} } = 0.1$ (illustrated by the
dashed-line in Fig~\ref{fig:TAimf}) which is a
reasonable approximation to the CMF of Taurus as observed by Onishi et
al. (2001).  The hashed histogram is the observed IMF of 
Taurus (Luhman et al. 2003a) and it compares well to the open histogram
given by our model.  The open circles show the contribution to the IMF
from ejected stars and brown dwarfs.  The binary fraction is very high
in our model as the vast majority of stars have formed in multiple
systems, only the ejected component has a low multiplicity.  This
again compares well with the high observed multiplicity in Taurus
(Duch\^{e}ne 1999).

This agrees well with the results of Goodwin et al. (2004c), the IMF is a
combination of a peak of bound systems with average stellar mass $\approx 1
M_\odot$ which remain bound in the cores, and a flat low-mass tail of
ejected brown dwarfs and low-mass stars.

\subsection{The IMFs of Orion and IC 348}

\begin{figure}
\centerline{\psfig{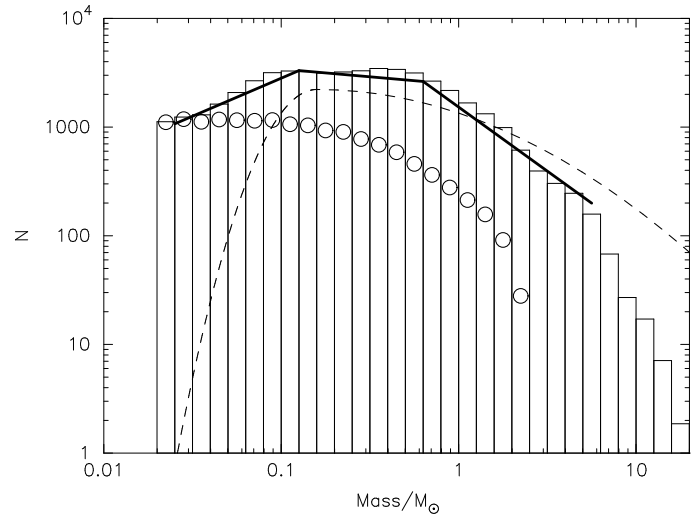}}
\caption{The IMF of Orion (solid line, from Muench et al. 2002) is
well-fitted by the open histogram produced by the CMF shown by the
dashed-line.  As in fig.~\ref{fig:TAimf}, the circles show the
contribution to the IMF from ejected stars.} 
\label{fig:Orion}
\end{figure}

Orion has an IMF that is very similar to the field (Muench et
al. 2002).  Fig.~\ref{fig:Orion} shows the fit to the Orion IMF given
by a CMF of mean $\bar{ {\rm log} M_{\rm core} } = -0.8$ and 
$\sigma_{ {\rm log} M_{\rm core} } = 0.3$ (lower) and $= 0.7$ (upper).

\begin{figure}
\centerline{\psfig{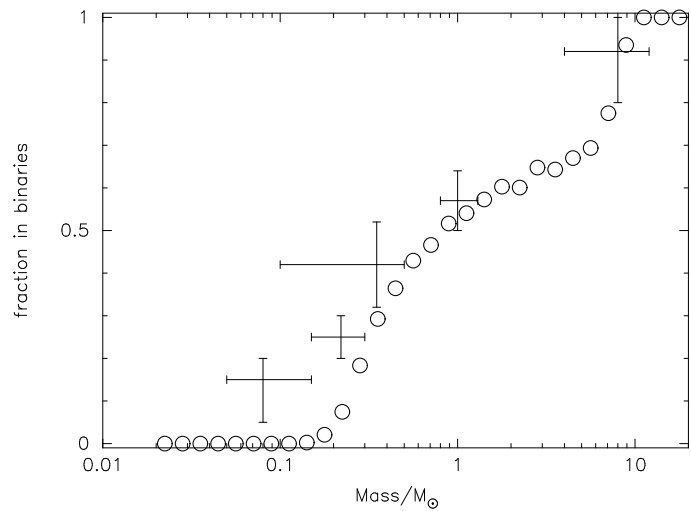}}
\caption{The binary fraction of stars in the model of Orion as a
function of primary mass (Open circles).  The error bars show the
observations of the field binary fraction adapted from Sterzik \&
Durisen (2003).}
\label{fig:Obin}
\end{figure}

Figure~\ref{fig:Orion} reproduces the IMF of Orion well with a wide,
flat peak between $0.1$ and $0.6 M_\odot$, falling at both ends, with
an approximately Salpeter slope at high-masses.  Fig.~\ref{fig:Obin}
shows the binary fraction as a function of primary mass.  This model fits
the observed field binary fractions quite well, except at lower masses where
it is assumed that all ejected stars and brown dwarfs are single
(which is not always the case, a low fraction of ejected stars are
multiples, see Goodwin et al. 2004b).

\begin{figure}
\centerline{\psfig{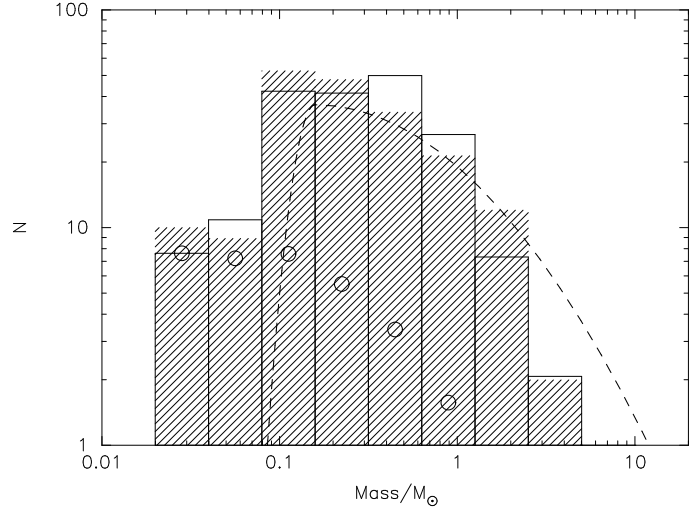}}
\caption{The IMF of IC 348 (hashed histogram, from Luhman et al. 2003b) is
well-fitted by the open histogram produced by the CMF shown by the
dashed-line.  As in fig.~\ref{fig:TAimf}, the circles show the
contribution to the IMF from ejected stars.} 
\label{fig:IC348}
\end{figure}

IC 348 has an IMF that is very similar to Orion except that it is
relatively deficient in brown dwarfs (Luhman et al. 2003b).
Fig.~\ref{fig:IC348} shows the fit to the IMF of IC 348 using a CMF of
mean $\bar{ {\rm log} M_{\rm core} } = -0.8$ and 
$\sigma_{ {\rm log} M_{\rm core} } = 0.1$ (lower) and $= 0.7$ (upper).
This is almost identical to the CMF used to model Orion, but the lower
extent of the CMF is far smaller (0.1 compared to 0.3 in the Orion 
CMF).  Almost no brown dwarfs are formed
in cores in IC 348, they are all the result of ejections from
higher-mass cores.

\section{Conclusions}

Using a simple model of fragmentation in cores we are able to match
the IMFs of  Taurus, NGC 2547, Orion and IC 348 with 
different core mass functions.  

The IMFs of Taurus and NGC 2547 are well-fitted with a CMF that peaks at a few 
solar masses, which matches the observed CMF of Taurus (Onishi et
al. 2002).  This CMF reproduces the peaks in these IMFs at $\sim 1
M_\odot$.  Most solar-type stars are formed in multiple systems, 
explaining the very high observed binary fraction in Taurus (Duch\^ene
1999).

To fit the IMFs of Orion and IC 348 requires CMFs that
peak at only a few tenths of a solar mass.  The lack of brown dwarfs
in IC 348 as compared to Orion can be explained by a CMF that does not
extend as far into the brown dwarf regime in IC 348.  The binary
fractions in Orion and IC 348 are close to those observed in the field.

\section*{Acknowledgements}

SPG is a UK Astrophysical Fluids Facility (UKAFF) Fellow.

\end{document}